\documentclass[10pt, a4paper, twocolumn]{IEEEtran}
\usepackage{amsmath,amsthm,epsfig,amssymb,verbatim,amsopn,cite,subfigure,multirow}
\usepackage{balance,nopageno}
\usepackage[usenames,dvipsnames]{color}
\usepackage[all]{xy}  %%% used to make block diagram
\usepackage{url}
\usepackage{amsfonts}
\usepackage{amssymb}
\usepackage{epsfig}
\usepackage{epstopdf}
\usepackage{balance}
\usepackage[margin=20mm,top=22mm,bottom=22mm]{geometry}%margins; tweak for printer deficiencies
  {\proof}{\proofend}
\newtheorem{proposition}{Proposition}

\DeclareMathOperator*{\argmax}{arg\,max}

\newcommand{\E}[1]{\mathbb{E}\left\{#1\right\}}

\newcommand{\qH}{{\bf H}}

\newcommand{\Ss}{\mathsf{S1}}
\newcommand{\Sss}{\mathsf{S2}}

\newcommand{\HSIkj}{h_{\mathsf{RR}}^{k,j}}
\newcommand{\GSIkj}{\gamma_{\mathsf{SI}}^{k,j}}
\newcommand{\GSIkkj}{\gamma_{\mathsf{SI}}^{k^*,j}}
\newcommand{\bGSI}{\bar{\gamma}_{\mathsf{SI}}}

\newcommand{\HSI}{\qH_{\mathsf{RR}}}

\newcommand{\HBRij}{h_{\mathsf{SR}}^{i,j}}
\newcommand{\GBRij}{\gamma_{\mathsf{SR}}^{i,j}}
\newcommand{\GBRiij}{\gamma_{\mathsf{SR}}^{i^*,j}}
\newcommand{\bGBR}{\bar{\gamma}_{\mathsf{SR}}}

\newcommand{\HBNu}{h_{\mathsf{SU1}}^{i}}
\newcommand{\GSNu}{\gamma_{\mathsf{SU1}}^{i}}
\newcommand{\bGSNu}{\bar{\gamma}_{\mathsf{SU1}}}

\newcommand{\HRNu}{h_{\mathsf{RU1}}^{k}}
\newcommand{\GRNu}{\gamma_{\mathsf{RU1}}^{k}}
\newcommand{\bGRNu}{\bar{\gamma}_{\mathsf{RU1}}}

\newcommand{\HRFu}{h_{\mathsf{RU2}}^{k}}
\newcommand{\GRFu}{\gamma_{\mathsf{RU2}}^{k}}
\newcommand{\GRFui}{\gamma_{\mathsf{RU2,AS}}}
\newcommand{\GRFus}{\gamma_{\mathsf{RU2,S1}}}

\newcommand{\GMR}{\gamma_{\mathsf{R}}}
\newcommand{\GMRi}{\gamma_{\mathsf{R,AS}}}
\newcommand{\GMRS}{\gamma_{\mathsf{R,S1}}}

\newcommand{\bGRFu}{\bar{\gamma}_{\mathsf{RU2}}}

\newcommand{\SRFu}{\sigma^2_{\mathsf{RU2}}}
\newcommand{\SRNu}{k_1\sigma^2_{\mathsf{RU1}}}
\newcommand{\SBNu}{\sigma^2_{\mathsf{SU1}}}
\newcommand{\SBR}{{\sigma}^2_{\mathsf{SR}}}

\newcommand{\Snuu}{\sigma_{n_{2}}^2}
\newcommand{\Sn}{\sigma_n^2}
\newcommand{\Sap}{\sigma_{\mathsf{SI}}^2}
\newcommand{\SXY}{\sigma_{\mathsf{XY}}^2}

\newcommand{\PS}{P_{\mathsf{S}}}
\newcommand{\PR}{P_{\mathsf{R}}}

\newcommand{\SUu}{\text{U1}}
\newcommand{\SUuu}{\text{U2}}
\newcommand{\SUui}{\text{U}u}

\newcommand{\Rly}{\mathsf{R}}
\newcommand{\Srs}{\mathsf{S}}
\newcommand{\Nu}{\mathsf{U1}}
\newcommand{\Fu}{\mathsf{U2}}
\newcommand{\Poutnu}{{\mathsf{P_{out,1}^{AS}}}}
\newcommand{\Poutfu}{{\mathsf{P_{out,2}^{AS}}}}

\newcommand{\PoutnuS}{{\mathsf{P_{out,1}^{S1}}}}
\newcommand{\PoutfuS}{{\mathsf{P_{out,2}^{S1}}}}
\newcommand{\PoutnuSS}{{\mathsf{P_{out,1}^{S2}}}}
\newcommand{\PoutfuSS}{{\mathsf{P_{out,2}^{S2}}}}

\newcommand{\MT}{M_{\mathsf{T}}}
\newcommand{\MR}{M_{\mathsf{R}}}
\newcommand{\MB}{N_{\mathsf{T}}}

\newcommand{\Prob}{\textnormal{Pr}}

\newcommand{\be}{\begin{equation}} \newcommand{\ee}{\end{equation}}
\newcommand{\bea}{\begin{eqnarray}} \newcommand{\eea}{\end{eqnarray}}

\usepackage{multibib}
\usepackage[nodisplayskipstretch]{setspace}
\newcites{Prim}{Very important papers}

\definecolor{light-gray}{gray}{0.65}

\newcounter{mytempeqcounter}

\title{\fontsize{0.88cm}{1cm}\selectfont Antenna Selection in Full-Duplex Cooperative NOMA Systems
}
\author{{Mohammadali Mohammadi$^\dag$, Zahra Mobini$^\dag$, Himal A. Suraweera$^\ddag$, and Zhiguo Ding$^\S$}\\
\small{$^\dag$Faculty of  Engineering, Shahrekord University, Iran\\
$^\ddag$Department of Electrical and Electronic Engineering, University of Peradeniya, Sri Lanka\\
$^\S$School of Computing and Communications, Lancaster University, United Kingdom \\
Email:  \{m.a.mohammadi, z.mobini\}@eng.sku.ac.ir,   himal@ee.pdn.ac.lk, z.ding@lancaster.ac.uk}}\normalsize

\begin{document}

\maketitle
\thispagestyle{empty}
\begin{abstract}
We investigate the problem of antenna selection (AS) in full-duplex (FD) cooperative non-orthogonal
multiple access (NOMA) systems, where a multi-antenna FD relay assists transmission from a multi-antenna base station (BS)
to a far user, while at the same, the BS transmits to a near user. Specifically, based on the end-to-end signal-to-interference-plus-noise ratio at the near and far users, two AS schemes to select a single
transmit antenna at both the BS and the relay, respectively, as well as a single receive antenna at relay are proposed.  In order to study the ergodic sum rate and outage probability of these AS schemes, we have derived closed-form expressions assuming Rayleigh fading channels.  The sum rate and outage probability of the AS schemes are also compared with the optimum selection scheme that maximizes the performance  as well as with a random AS scheme. Our results show that the proposed AS schemes can deliver a near-optimal performance for near and far users, respectively.
\end{abstract}
%***************************************************************************
%%***************************************************************************
%\newpage
\section{Introduction}
%***************************************************************************
%%***************************************************************************
Each new generation of wireless communication systems has been designed to support the demands of increased traffic and data throughput with spectral efficiency as a key factor. To this end, non-orthogonal multiple access (NOMA) principle has been recognized as a key technology that can improve the spectral efficiency of 5G wireless systems. In contrast to traditional orthogonal multiple access (OMA) techniques, NOMA multiplexes signals between users with strong/weak channel conditions (a.k.a. near/far users) in the power domain and apply successive interference cancellation (SIC) at the receivers to remove the inter-user interference~\cite{Saito:VTC2013,Ding:Survay,Zhiguo:CLET:2015}.

On the other hand, multiple-input multiple-output (MIMO) technology that offers substantial performance gains has become an integral part of modern communication systems. However, the benefits of MIMO come at the price of increased computational complexity and cost of hardware radio frequency chains that scales with the number of antennas~\cite{Molisch}. In this context, antenna selection (AS) schemes with low implementation complexity and perform close to traditional MIMO systems have been touted as a practical solution in the literature.  There is a sizable matured body of work on AS for different MIMO systems. To this end, AS in combination with NOMA has received interest in the recent literature~\cite{Vucetic:ICC17,Costa:ICC2017,Serpedin:TVT:2017}, however the topic is still in infancy. Especially, the complexity of deciding on AS solutions in NOMA systems are exacerbated due to the complicated nature of near/far user performance criterion~\cite{Vucetic:ICC17}.

In this paper, we analyze the AS performance of a full-duplex (FD) cooperative NOMA system. FD is another promising technology considered for 5G implementation. Furthermore, the marriage between FD and NOMA can boost the performance as confirmed in~\cite{Caijun:CLET:2016,Zahra:GC17,
Yue:tcom:2017,Mohammadi:GC17} so far. The main bottleneck for FD operation is the self-interference (SI)~\cite{Choi:mobicom:2010,Riihonen:JSP:2011,Sabharwal:TWC2012}. Thus, AS should consider the effect of SI during the selection of strong channels toward the near/far users. This makes the AS problem in FD NOMA systems a much more complicated affair than in half-duplex (HD) NOMA systems~\cite{Vucetic:ICC17}. Specifically, for the considered FD NOMA relay system, we propose AS schemes to achieve near/far user end-to-end (\emph{e2e}) signal-to-interference noise ratio (SINR) maximization and study the sum rate and outage probability, respectively. Our contributions can be summarized as follows:

\begin{itemize}
\item Two low complexity AS schemes, i.e., max-$\SUu$ AS scheme and max-$\SUuu$ AS scheme are proposed to maximize the \emph{e2e} SINR at the near and far user, respectively.  The performance of the FD cooperative NOMA system with the two AS schemes is analyzed by deriving exact ergodic sum rate and outage probability expressions.
\item Our findings reveal that max-$\SUu$ AS scheme can significantly improve the system sum-rate, while max-$\SUuu$ AS scheme can provide better user fairness. In particular, max-$\SUu$ AS scheme can achieve near optimum rate performance in the entire SNR range.
\end{itemize}

\emph{Notation:} We use $\mathbb{E}\left\{X\right\}$ to denote the expected value of the random variable (RV) $X$; its  probability density function (pdf) and  cumulative distribution function (cdf) are $f_X(\cdot)$ and $F_X(\cdot)$  respectively; $\mathcal{CN}(\mu,\sigma^2)$ denotes a circularly symmetric complex Gaussian RV $X$ with mean $\mu$ and variance $\sigma^2$ and $\mathrm{E_i}(x)=\int_{-\infty}^{x}\frac{e^t}{t}dt$ is the exponential integral function~\cite[Eq. (8.211.1)]{Integral:Series:Ryzhik:1992}.

%%%%%%%%%%%%%%%%%%%%%%%%%%%%%%%%%%%%%%%%%%%%%%%%%%%%%%%%%%%%%%%%
\section{System Model}\label{sec:system model}
\begin{figure}[htb!]
\centering
\includegraphics[width=92mm, height=160mm]{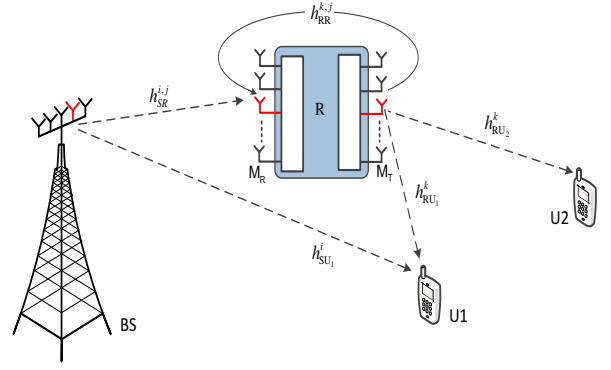}
\vspace{-31em}
\caption{ FD Cooperative NOMA system with antenna selection.}\label{fig:model}
\end{figure}

%%%%%%%%%%%%%%%%%%%%%%%%%%%%%%%%%%%%%%%%%%%%%%%%%%%%%%%%%%%%%%%%
Consider a two user NOMA downlink system where $\SUu$ (near user) communicates directly with the base station (BS), while $\SUuu$ (far user) requires the assistance of a multi-antenna FD relay, $\Rly$ as shown in Fig.~\ref{fig:model}. Both $\SUu$ and $\SUuu$ are equipped with a single antenna each, the BS is equipped with $\MB$ antennas, while $\Rly$ is equipped with two group of antennas: $\MR$ receive antennas and $\MT$ transmit antennas.

In order to keep the implementation complexity low, we assume that the BS and $\Rly$ perform single AS~\cite{Molisch}. To be precise, the BS selects one (e.g. $i$-th) out of $\MB$ available transmit antennas, while $\Rly$ selects one (e.g. $j$-th)  out of $\MR$ available antennas to receive signals. Moreover, one antenna (e.g. $k$-th) out of $\MT$ transmit antennas is selected at $\Rly$ to forward the BS signal to $\SUuu$.

We assume that all channels experience independent Rayleigh fading and that they remain constant over one transmission slot. The channel between the $j$-th receive and the $i$-th transmit antenna from terminal $X$ to terminal $Y$, is denoted by $h_{XY}^{i,j}\sim\mathcal{CN}(0,\SXY)$ where $X \in\{\Srs,\Rly\}$ and $Y \in\{\Rly, \Nu,\Fu\}$.
%If the receiver has single antenna, the index $j$ will be omitted, i.e., $h_{XY}^{i,1} = h_{XY}^{i}$.

%============================================
%=================================
\subsection{Transmission Protocol}
%=================================
According to the NOMA concept~\cite{Ding:Survay,Zhiguo:CLET:2015}, the BS transmits a combination of intended messages to both users as
%=======================================
\vspace{-0.2em}
\begin{align}
s[n]=\sqrt{\PS a_1}x_1[n]+\sqrt{\PS a_2}x_2[n],
\end{align}
%=======================================
where $x_i$, $i\in\{1,2\}$ denotes the information symbol intended for $\text{U}i$, and $a_i$ denotes the power allocation coefficient,
such that $a_1 + a_2 =1$ and $a_1< a_2$.

Since $\Rly$ is FD capable, it transmits the decoded symbol $x_2[n-\tau]$, where $\tau\geq1$ accounts for the time delay caused by FD processing at $\Rly$~\cite{Riihonen:JSP:2011}. Therefore, the receive signal at $\Rly$ can be written as~\cite{Caijun:CLET:2016}
%=======================================
\vspace{-0.0em}
\begin{align}
y_R[n]&= \HBRij s[n]+\sqrt{\PR}\HSIkj x_2[n-\tau]+ n_R[n],
\end{align}
%=======================================
where $\PR$ is the relay transmit power and $n_R[n]\sim\mathcal{CN}(0,\Sn)$ is the additive white Gaussian noise (AWGN) at $\Rly$. We assume imperfect SI cancellation due to FD operation at $\Rly$ and model the elements of the  $\MR\times\MT$ residual SI channel $\HSI =[\HSIkj]$ as independent identically distributed (i.i.d) $\mathcal{CN}(0,\Sap)$ RVs~\cite{Riihonen:JSP:2011}.

The information intended for $\SUuu$ is decoded at  $\Rly$ with SIC treating the symbol of $\SUu$ as interference~\cite{Caijun:CLET:2016}. Hence, the SINR at $\Rly$ can be written as
%=====================================
\vspace{-0.4em}
\begin{align}\label{eq:SINR relay}
&\hspace{-0.2em}
\GMR=\frac{ a_2\GBRij}{a_1\GBRij \!+\!\! \GSIkj\!+1},
\end{align}
%=====================================
where $\GBRij=\rho_S|\HBRij|^2$ and $\GSIkj=\rho_R|\HSIkj|^2$ with $\rho_S=\frac{\PS}{\Sn}$ and $\rho_R=\frac{\PR}{\Sn}$.

At the same time, $\SUu$ receives the following signal
%=======================================
\vspace{-0.0em}
\begin{align}\label{eq:sig:U1}
y_1[n]\!=\!\HBNu s[n]\!+\!\sqrt{\PR} \HRNu x_2[n\!-\!\tau]\!+\!n_1[n],
\end{align}
%=======================================
where $n_1[n]\sim\mathcal{CN}(0,\Sn)$ is the AWGN at $\SUu$.

%%%%$$$$$$$$$$$$$$$$$$$$$$$$$$$$$$$$$$$$$$$$$$$$$$$$$$$$$$$$$$$$$$$$$$$$$$$$$$$$$$$$$$$$$$$$$$$$$$$$$$$$$$$$$$$$$$$$$$$$$$$$$
%%%%$$$$$$$$$$$$$$$$$$$$$$$$$$$$$$$$$$$$$$$$$$$$$$$$$$$$$$$$$$$$$$$$$$$$$$$$$$$$$$$$$$$$$$$$$$$$$$$$$$$$$$$$$$$$$$$$$$$$$$$$$

Based on~\eqref{eq:sig:U1}, the SINR of $\SUuu$ observed at $\SUu$ can be written as
%=======================================
\vspace{-0.0em}
\begin{align}\label{eq:SINR UE2 at UE1}
\gamma_{12}&=\frac{ a_2\GSNu}{ a_1\GSNu +\GRNu+1},
\end{align}
%=======================================
where $\GSNu = \rho_S|\HBNu|^2$ and  $\GRNu=\rho_R|\HRNu|^2$.

It is assumed that the symbol, $x_{2}[n-\tau]$, is priory known to $\SUu$ and thus $\SUu$ can be removed~\cite{Caijun:CLET:2016}. However, by considering realistic imperfect interference cancellation wherein $\SUu$ cannot perfectly remove $x_{2}[n-\tau]$, we model $\HRNu \sim \mathcal{CN}(0,k_1\sigma^2_{\mathsf{RU1}})$ as the inter-user interference channel where the parameter $k_1$ presents the strength of inter-user interference~\cite{Caijun:CLET:2016}. If $\SUu$ perfectly cancels the $\SUuu$'s signal, the SINR at $\SUu$ is given by
%=======================================
\vspace{-0.0em}
\begin{align}\label{eq:SINR at UE1}
\gamma_{1}&=\frac{ a_1\GSNu}{\GRNu+1}.
\end{align}
%=======================================

Moreover, the received signal at $\SUuu$, from $\Rly$ can be written as
%===========================
\vspace{-0.4em}
\begin{align}
y_2[n]=\sqrt{\PR}  \HRFu x_2[n-\tau]+n_2[n],
\end{align}
where $n_2[n]\sim\mathcal{CN}(0,\Snuu)$ denotes the AWGN at $\SUuu$. Hence, the signal-to-noise ratio (SNR) at $\SUuu$ is given by
%===========================
\vspace{-0.4em}
\begin{align}\label{eq:gamR2}
\GRFu=\frac{\PR}{\Snuu }|\HRFu|^2.
\end{align}
%===========================
The \emph{e2e} SINR at $\SUuu$ can be expressed as
\begin{align}\label{eq:e-2-e far}
\hspace{-0.3em}\gamma_{2}&\!=\!\min\!\left(\!\frac{ a_2\GSNu}{ a_1\GSNu\!+\!\GRNu\!+\!1},\frac{a_2\GBRij}{ a_1\GBRij\!+\!\GSIkj\!+\!1},
\GRFu\!\right)\!.
\end{align}

%%%%%%%%%%%%%%%%%%%%%%%%%%%%%%%%%%%%%%%%%%%%%%%%%
%\section{Problem Formulation}
\subsection{Antenna Selection Schemes}
%%%%%%%%%%%%%%%%%%%%%%%%%%%%%%%%%%%%%%%%%%%%%%%%%
We now propose two AS schemes for the described FD cooperative NOMA system where joint selection of single transmit and receive antennas at the BS and $\Rly$ based on \emph{e2e} SINRs at the near user, $\SUu$ and far user, $\SUuu$ is performed.

%%%%%%%%%%%%%%%%%%%%%%%%%%%%%%%%%%%%%%%
\subsubsection{max-$\SUu$ AS Scheme}
%%%%%%%%%%%%%%%%%%%%%%%%%%%%%%%%%%%%%%%
This scheme selects antennas first to  maximizes the \emph{e2e} SINR at $\SUu$,~\eqref{eq:SINR at UE1} and next with remaining AS choices tries to maximize also \emph{e2e} SINR at U2. Therefore, the particular AS scheme can be mathematically expressed as
%====================================
\begin{align}\label{eq:ASc near}
\{i^*,  k^*\} &=\argmax_{\substack{1\leq i\leq\MB,  1\leq k\leq\MT}}\frac{ a_1\GSNu}{\GRNu+1}\nonumber\\
j^*&=\argmax_{\substack{ 1\leq j\leq\MR}}\frac{a_2\GBRiij}{ a_1\GBRiij + \GSIkkj+1}.
\end{align}
%=======================================
\subsubsection{max-$\SUuu$ AS Scheme}
The max-$\SUuu$ AS scheme achieves \emph{e2e} SINR maximization at $\SUuu$ according to
%=================================
\begin{align}\label{eq:ASc far}
\{i^*, j^*, k^*\} &\!= \!\!\!\argmax_{\substack{1\leq i\leq\MB, 1\leq j\leq\MR,\\ 1\leq k\leq\MT}}
\!\!\!\!\!\min\left(\frac{ a_2\GSNu}{ a_1\GSNu \!+\GRNu\!+1},\right.\nonumber\\
&\left.\hspace{3em}\frac{a_2\GBRij}{ a_1\GBRij + \GSIkj+1},
\GRFu\right).
\end{align}
%=========================

According to~\eqref{eq:ASc far} there is no degree-of-freedom (in terms of AS) available
for maximizing the \emph{e2e} SINR at $\SUu$. Therefore,  in terms of the near user SINR, max-$\SUuu$ AS scheme is inferior to that of the max-$\SUu$ AS scheme.

It worth pointing out that other AS schemes, for example with criterion such as to maximize the near/far user performance subject to a pre-defined far/near performance are also possible to design. Analyzing such AS schemes whose performance remain in between  max-$\SUu$ and max-$\SUuu$ AS schemes is relegated to the journal version of this work.

%%%%%%%%%%%%%%%%%%%%%%%%%%%%%%%%%%%%%%%%%%%%%%%%%%%%%%%%%%
%\subsection{Implementation and Signal Requirements}
%%%%%%%%%%%%%%%%%%%%%%%%%%%%%%%%%%%%%%%%%%%%%%%%%%%%%%%%%%
In order to implement max-$\SUu$ and max-$\SUuu$ AS schemes, the BS can transmit a pilot signal and $\Rly$ can decide on the best antenna indexes to be used at the BS and its receive side. $\Rly$ can also transmit a pilot signal to $\SUu$ and $\SUuu$ from each of the relay antenna. Upon reception of the pilot, $\SUu$ (in max-$\SUu$ AS scheme) and $\SUuu$ (in max-$\SUuu$ AS scheme) can next feedback the antenna index that $\Rly$ must used in subsequent information transmission. Moreover, in max-$\SUu$ AS scheme, upon reception of the BS pilot signal, $\SUu$ can decide the best antenna and feedback the corresponding index to the BS to commence information transmission.
%%%%%%%%%%%%%%%%%%%%%%%%%%%%%%%%%%%%%%
\vspace{-0.5em}
\section{Performance Analysis}
%%%%%%%%%%%%%%%%%%%%%%%%%%%%%%%%%%%%%%
In this section, we investigate the performance of proposed AS schemes in terms of the
achievable ergodic sum rate and outage probability. The derived results will enable us to examine the
benefits of the proposed AS schemes.
%%%%%%%%%%%%%%%%%%%%%%%%%%%%%%%%%%%%%%%%%%%%%%%%%%%%
\vspace{-0.2em}
\subsection{Ergodic Sum Rate}\label{subsec:rate}
%%%%%%%%%%%%%%%%%%%%%%%%%%%%%%%%%%%%%%%%%%%%%%%%%%%%
The ergodic achievable sum rate of the FD cooperative NOMA system is given by
%==============================
\vspace{-0.2em}
\begin{align}\label{eq:sum rate}
\mathcal{R}_{sum} =\mathcal{R}_{\SUu}^{\mathsf{AS}} +\mathcal{R}_{\SUuu}^{\mathsf{AS}},
\end{align}
%=============================
with
%=============================
\begin{subequations}
\begin{align}
\mathcal{R}_{\SUu}^{\mathsf{AS}} &= \mathbb{E}\left\{\log_2(1+\gamma_{1,\mathsf{AS}})\right\},\\
\mathcal{R}_{\SUuu}^{\mathsf{AS}}&= \mathbb{E}\left\{\log_2(1+\gamma_{2,\mathsf{AS}})\right\},
\end{align}
\end{subequations}
%=============================
where $\gamma_{1,\mathsf{AS}}$ and $\gamma_{2,\mathsf{AS}}$ denote the \emph{e2e} SINR at the $\SUu$ and $\SUuu$, corresponding to the specific AS scheme with $\mathsf{AS}\in\{\Ss, \Sss\}$. We use $\Ss$ to refer to the max-$\SUu$ AS scheme and $\Sss$ to refer to the max-$\SUuu$ AS scheme.

Before we proceed, it is useful, to note that for a nonnegative RV $X$, since $\E{X} = \int_{t=0}^{\infty}\Prob(X>t)dt$,  the ergodic achievable rate for near user ($u=1$) or far user ($u=2$) can be written as
%=================================================
\begin{align}\label{eq:rate and cdf}
\mathcal{R}_{\SUui}^{\mathsf{AS}}=
\frac{1}{\mathrm{ln} 2}
\int_{0}^{\infty}\frac{1-F_{\gamma_{u,\mathsf{AS}}}(x)}{1+x}dx,
\end{align}
%=============================
where $F_{\gamma_{u,\mathsf{AS}}}(z) =\Prob( \gamma_{u,\mathsf{AS}}\leq z)$ is the cdf of the  RV $\gamma_{u,\mathsf{AS}}$.

In the sequel, we present key results for the near and far user ergodic rates due to the proposed AS schemes.

\begin{proposition}\label{prop:rate:S1}
The ergodic achievable rates of $\SUu$ and $\SUuu$ of max-$\SUu$ AS scheme, are respectively given by
%==============================
\begin{align}\label{eq:R1:S1}
 &\mathcal{R}_{\SUu}^{\Ss}= \frac{\MB}{\mathrm{ln} 2}\sum_{p=0}^{\MB-1}
\frac{ (-1)^p\binom{\MB-1}{p} }
{(p+1)\left( \frac{(p+1)\bGRNu }{\MT a_1\bGSNu}-1\right)}\\
&\times\left( e^{\frac{1}{\bGRNu}}\mathrm{E_i}\left(-\frac{1}{\bGRNu}\right)-e^{\frac{(p+1)}{a_1\bGSNu}}\mathrm{E_i}
\left(-\frac{(p+1)}{a_1\bGSNu}\right)\right)\nonumber
\end{align}
%==============================
and
\begin{align}\label{eq:R2:S1}
&\mathcal{R}_{\SUuu}^{\Ss}= \frac{\MR\MB}{\mathrm{ln} 2}
\int_{0}^{\infty}
\frac{e^{-\frac{x}{\bGRFu}}}{1+x}\nonumber\\
&\hspace{2em}\times\sum_{p=0}^{\MB-1}\frac{(-1)^p\binom{\MB-1}{p}e^{-\frac{(p+1)x}{\bGSNu\left(a_2-a_1x\right)}}}
{(p+1)\left(1+\frac{\bGRNu}{\MT\bGSNu}\frac{(p+1)x}{ \left(a_2-a_1x\right)}\right)}\nonumber\\
&\hspace{2em}\times\sum_{q=0}^{\MR-1}\frac{(-1)^q\binom{\MR-1}{q}e^{-\frac{(q+1)x}{\bGBR\left(a_2-a_1x\right)}}}
{(q+1)\left(1+\frac{\bGSI}{\bGBR}\frac{(q+1)x}{ (a_2-a_1x)}\right)} dx,
\end{align}
%=============================
where  $\bGBR = \rho_S\SBR$, $\bGSNu=\rho_S\SBNu$, $\bGRNu=\rho_R\SRNu$, $\bGRFu=\rho_R\SRFu$ and $\bGSI=\rho_R \Sap$.
\end{proposition}

\emph{Proof:} According to~\eqref{eq:rate and cdf}, the ergodic achievable rate can be calculated from the cdf of~\eqref{eq:SINR at UE1} and~\eqref{eq:e-2-e far}. Let us start with the ergodic achievable rate of $\SUu$. By invoking~\eqref{eq:ASc near}, the ratio $\frac{a_1\GSNu}{\GRNu+1}$ is maximized when the strongest BS-$\SUu$ channel and the weakest $\Rly$-$\SUu$ channel are selected. Therefore, the cdf of $\gamma_{1,\Ss}$ can be
evaluated as
%==============================
\begin{align*}
F_{\gamma_{1,\Ss}}(x)=\int_{0}^{\infty} F_A\left((y+1)x/a_1\right)f_B(y)dy,
 \end{align*}
%=============================
 where $A$ is a RV defined as the maximum out of $\MB$ exponentially distributed independent RVs, while $B$ is the minimum out of $\MT$ exponentially distributed independent RVs. Substituting the required cdf and the pdf and simplifying yields
%==============================
\begin{align}\label{eq:cdf:gam1:S1}
F_{\gamma_{1,\Ss}}(x) &\!=\! 1\!- \!\MB\!\!\sum_{p=0}^{\MB-1}\!\!
\frac{ (-1)^p\binom{\MB-1}{p} e^{-\frac{(p+1)x}{a_1\bGSNu}}}
{(p+1)\left(1 \!+\! \frac{(p+1)\bGRNu x}{\MT a_1\bGSNu}\right)}\!.
\end{align}
%=============================
Next, substituting~\eqref{eq:cdf:gam1:S1} into~\eqref{eq:rate and cdf}, then with the help of the integration identity in~\cite[Eq. (3.352.4)]{Integral:Series:Ryzhik:1992}, and after some algebraic manipulations, we arrive at~\eqref{eq:R1:S1}.

We now turn our attention to evaluate $\mathcal{R}_{\SUuu}^{\Ss}$. To this end, the cdf of $\gamma_{2,\Ss}$ can be expressed as
%==============================
\begin{align}\label{eq:cdf:gam2:def}
F_{\gamma_{2,\Ss}}(x) &= \Prob\left(\min\left(\gamma_{12,\Ss},\GMRS,\GRFus\right)<x\right)\nonumber\\
&=1 - \Prob\left(\gamma_{12,\Ss}>x\right)\Prob\left(\GMRS>x\right)\nonumber\\
&\hspace{3em}\times\Prob\left(\GRFus>x\right).
\end{align}
%=============================

According to the criterion in~\eqref{eq:ASc near}, for $x<\frac{a_2}{a_1}$, we can write
%=========================
\begin{align}\label{eq:cdf:gam12S1}
\Prob\left(\gamma_{12,\Ss}\!>\!x\right)
%& = \Prob\left(\frac{ a_2\GSNu}{ a_1\GSNu +\GRNu+1}>x\right)\nonumber\\
%&\hspace{-5em} = \Prob\left(\GSNu >\frac{\GRNu+1}{\frac{a_2}{x}-a_1}\right)\nonumber\\
&=\!1\!-\!\int_{0}^{\infty}\!\!F_{\GSNu}\left(\frac{(y+1)x}{a_2\!-a_1x}\right)f_{\GRNu}(y)dy\nonumber\\
&\hspace{-4.5em}=\frac{\MB\MT}{\bGRNu}\sum_{p=0}^{\MB-1}\frac{(-1)^p\binom{\MB-1}{p}}{p+1} e^{-\frac{(p+1)x}
{\bGSNu\left(a_2-a_1 x\right)}}\nonumber\\
&\hspace{-0em}\times\int_{0}^{\infty} e^{-\frac{(p+1)yx}{\bGSNu\left(a_2-a_1x\right)}} e^{-\frac{\MT y}{\bGRNu}}dy\nonumber\\
&\hspace{-4.5em}=\MB\sum_{p=0}^{\MB-1}\frac{(-1)^p\binom{\MB-1}{p}e^{-\frac{(p+1)x}{\bGSNu\left(a_2-a_1x\right)}}}
{(p+1)\left(1+\frac{\bGRNu}{\bGSNu}\frac{(p+1)x}{\MT \left(a_2-a_1x\right)}\right)},
\end{align}
%========================
where the second equality follows since $f_{\GRNu}(y)=\frac{\MT}{\bGRNu}e^{-\frac{\MT y}{\bGRNu}}$ and $F_{\GSNu}(x) = (1-e^{-\frac{x}{\bGSNu}})^{\MT}$ can be written as
%================================
\begin{align}\label{eq:cdf:GSNu}
F_{\GSNu}(x) = 1-\MB\sum_{p=0}^{\MB-1}\frac{(-1)^p\binom{\MB-1}{p}}{p+1} e^{-\frac{(p+1)x}{\bGSNu}}.
\end{align}
%================================

Moreover, based on~\eqref{eq:ASc near}, for the selected transmit antennas at the BS and $\Rly$, the ratio $\frac{a_2\GBRiij}{ a_1\GBRiij + \GSIkkj+1}$ can be maximized when the strongest BS-$\Rly$ channel and weakest SI channel are selected. However, theses two channels are coupled with each other through the selected antenna at the $\Rly$ input. Therefore, it is difficult, if not impossible, to find the cdf of $\frac{a_2\GBRiij}{ a_1\GBRiij + \GSIkkj+1}$. Alternatively, we propose to select the receive antenna at $\Rly$ such that $\GBRiij$ is maximized, which will be shown to be a good approximation across the entire SNR range in Section~\ref{sec:num} (cf. Optimum AS (U2)). Therefore,  we have
%================================
\begin{align*}
\Prob\left(\GMRS>x\right)=1-\int_{0}^{\infty} F_A\left(\frac{(y+1)x}{a_2-a_1x}\right)f_B(y)dy,
\end{align*}
%================================
for $x<\frac{a_2}{a_1}$, where $A$ is a RV defined as the largest out of $\MR$ exponentially distributed independent RVs, and since SI link is ignored, $B$ is an exponentially distributed RV with parameter $\bGSI$. Substituting the required cdf and the pdf and
simplifying yields
%=========================
\begin{align}\label{eq:cdf:GMRS}
\Prob\left(\GMRS\!>\!x\right)&\! =\! \MR\!\!\!\sum_{q=0}^{\MR-1}\!\!\frac{(-1)^q\binom{\MR-1}{q}e^{-\frac{(q+1)x}{\bGBR\left(a_2-a_1 x\right)}}}
{(q\!+\!1)\left(1\!+\!\frac{\bGSI}{\bGBR}\frac{(q+1)x}{ \left(a_2-a_1 x\right)}\right)}\!.
\end{align}
%========================

Finally, since the $\Rly$-$\SUuu$ link is ignored, we have
$\Prob(\GRFus>x) = e^{-\frac{x}{\bGRFu}}$.  To this end, pulling everything together, we obtain
%==============================
\begin{align}\label{eq:cdf:gam2:S1}
F_{\gamma_{2,\Ss}}(x) &= 1 - \MR\MB e^{-\frac{x}{\bGRFu}}\nonumber\\
&\hspace{0.5em}\times\sum_{p=0}^{\MB-1}\frac{(-1)^p\binom{\MB-1}{p}e^{-\frac{(p+1)}{\bGSNu\left(\frac{a_2}{x}-a_1\right)}}}
{(p+1)\left(1+\frac{\bGRNu}{\bGSNu}\frac{(p+1)}{\MT \left(\frac{a_2}{x}-a_1\right)}\right)}\nonumber\\
&\hspace{0.5em}\times\sum_{q=0}^{\MR-1}\frac{(-1)^q\binom{\MR-1}{q}e^{-\frac{(q+1)x}{\bGBR\left(a_2-a_1x\right)}}}
{(q+1)\left(1+\frac{\bGSI}{\bGBR}\frac{(q+1)x}{ \left(a_2-a_1 x\right)}\right)}.
\end{align}
%=============================

Having obtained the cdf of $\gamma_{2,\Ss}$,  the ergodic rate of $\SUuu$ can be obtained by employing~\eqref{eq:rate and cdf}. For $x>\frac{a_2}{a_1}$ it can be readily checked that $F_{\gamma_{2,\Ss}}(x)= 1$ and hence $\mathcal{R}_{\SUuu}^{\Ss}=0$.
$\blacksquare$

\begin{proposition}\label{prop:rate:S2}
The ergodic achievable rates of $\SUu$ and $\SUuu$ of max-$\SUuu$ AS scheme, are respectively given by
%==============================
\begin{align}\label{eq:R1:S2}
\mathcal{R}_{\SUu}^{\Sss}&=\frac{1}{\mathrm{ln} 2}
\frac{a_1\bGSNu}{\left(\bGRNu-a_1\bGSNu\right)}\\
&\hspace{1em}\times\left( e^{\frac{1}{\bGRNu}}\mathrm{E_i}\left(\!-\frac{1}{\bGRNu}\right)\!-e^{\frac{1}{a_1\bGSNu}}\mathrm{E_i}
\left(\!-\frac{1}{a_1\bGSNu}\right)\right)\!,\nonumber
\end{align}
and
\begin{align}\label{eq:R2:S2}
&\mathcal{R}_{\SUuu}^{\Sss}= \frac{\MT\MB}{\mathrm{ln} 2}
\int_{0}^{\infty}\frac{e^{-\frac{ x}{\bGSNu\left(a_2-a_1x\right)}}}
{\left(1 + \frac{\bGRNu}{\bGSNu}\frac{ x}{\left(a_2-a_1x\right)}\right)(1+x)}\nonumber\\
&\hspace{6.5em}\times\sum_{p=0}^{\MB-1}\frac{(-1)^p\binom{\MB-1}{p}e^{-\frac{(p+1)x}{\bGBR\left(a_2-a_1x\right)}}}
{(p+1)\left(1+\frac{\bGSI}{\MR\bGBR}\frac{(p+1)x}{ (a_2-a_1x)}\right)}\nonumber\\
&\hspace{6.5em}\times\sum_{q=0}^{\MT-1}
\frac{ (-1)^q\binom{\MT-1}{q} e^{-\frac{(q+1)x}{\bGRFu}}}
{(q+1)}dx.
\end{align}
%=============================
\end{proposition}
%==============================

%==============================
\emph{Proof:} The proof follows similar steps to the proof of Proposition~\ref{prop:rate:S1} and thus only an outline is presented.
According to~\eqref{eq:ASc far} the \emph{e2e} SINR at $\SUuu$ is maximized when each term inside in the minimum function is maximized. Therefore, a transmit antenna at $\Rly$ is selected  such that $\GRFu$ is maximized, i.e., $\GRFu$ is the
largest of  $\MT$ exponential RVs with parameter $\bGRFu$. Moreover, since SI is the main source of performance degradation in FD mode, for a particular transmit antenna at $\Rly$, the best receive antenna at the $\Rly$ is selected such that the SI strength is minimized~\cite{Suraweera:TWC:2014}. Hence, $\GSIkkj$ is the minimum of $\MR$ exponential RVs with parameter $\bGSI$. Finally, for given $k^*$ and $j^{*}$, a best transmit antenna at the BS is selected such that $\gamma_{\mathsf{R}}$ is maximized. Hence, a single transmit antenna at the BS is selected such that the SINR at BS-$\Rly$ is maximized for the $i^*$-th receive antenna at $\Rly$.

Collecting the required cdfs and  pdfs and simplifying yields
%====================================
\vspace{-0.2em}
\begin{align}\label{eq:cdf:gam1:S2}
F_{\gamma_{1,\Sss}}(x) &= 1- \frac{e^{-\frac{x}{a_1\bGSNu}}}{1+\frac{\bGRNu}{a_1\bGSNu}x},
\end{align}
%=============================
and
%==============================
\begin{align}\label{eq:cdf:gam2:S2}
F_{\gamma_{2,\Sss}}(x) & = 1-\MT\MB\frac{e^{-\frac{x }{\bGSNu\left(a_2-a_1x\right)}}}
{1 + \frac{\bGRNu}{\bGSNu}\frac{x}{\left(a_2-a_1x\right)}}\nonumber\\
&\times\sum_{p=0}^{\MB-1}\frac{(-1)^p\binom{\MB-1}{p}e^{-\frac{(p+1)c}{\bGBR\left(a_2-a_1x\right)}}}
{(p+1)\left(1+\frac{\bGSI}{\bGBR}\frac{(p+1)x}{\MR \left(a_2-a_1x\right)}\right)}\nonumber\\
&\times\sum_{q=0}^{\MT-1}
\frac{ (-1)^q\binom{\MT-1}{q} e^{-\frac{(q+1)x}{\bGRFu}}}
{(q+1)},
\end{align}
%=============================
respectively. To this end, by invoking~\eqref{eq:rate and cdf},~\eqref{eq:cdf:gam1:S2}, and~\eqref{eq:cdf:gam2:S2} after some algebraic manipulations, we arrive at the desired result.
$\hspace{17.5em}\blacksquare$

%%%%%%%%%%%%%%%%%%%%%%%%%%%%%%%%%%%%%%
\subsection{Outage Probability}
%%%%%%%%%%%%%%%%%%%%%%%%%%%%%%%%%%%%%%
Outage probability is a key metric used to measure  the event that the data rate supported by instantaneous channel realizations is
less than a targeted user rate. Therefore, the outage probability is an important performance metric to characterizes the performance of NOMA systems~\cite{Zhiguo:CLET:2015}.

The following Propositions present exact closed-form expressions for the outage probability of max-$\SUu$ and max-$\SUuu$ AS schemes
\begin{proposition}
The outage probability of $\SUu$ with max-$\SUu$ and max-$\SUuu$ AS scheme, are
respectively given by
%===========================
%\begin{subequations}
\begin{align}\label{eq:outNEARS1}
&\PoutnuS\!=\!1\!- \!\MB\!\!\sum_{p=0}^{\MB-1}\!\!
\frac{ (-1)^p\binom{\MB-1}{p} e^{-\frac{(p+1)\zeta}{\bGSNu}}}
{(p+1)\left(1 \!+\! \frac{\bGRNu}{\bGSNu}\frac{(p+1) \zeta}{\MT}\right)},
\end{align}
and
\begin{align}\label{eq:outNEARS2}
&\PoutnuSS
=1- \frac{e^{-\frac{\zeta}{\bGSNu}}}{1+\frac{\bGRNu}{\bGSNu}\zeta},
\end{align}
%\end{subequations}
%===========================
where $\zeta={ \max\left(\frac{\theta_2}{a_2-a_1\theta_2},\frac{\theta_1}{a_1} \right)}$, $\theta_1 = 2^{\mathcal{R}_1}-1$ and $\theta_2 = 2^{\mathcal{R}_2}-1$  with $\mathcal{R}_1$ and $\mathcal{R}_2$ being the transmission rates at $\SUu$ and $\SUuu$.
\end{proposition}

%%%%$$$$$$$$$$$$$$$$$$$$$$$$$$$$$$$$$$$$$$$$$$$$$$$$$$$$$$$$$$$$$$$$$$$$$$$$$$$$$$$$$$$$$$$$$$$$$$$$$$$$$$$$$$$$$$$$$$$$$$$$$
%\vspace{-2em}
\begin{figure}[htb!]
\centering
\includegraphics[width=85mm, height=66mm]{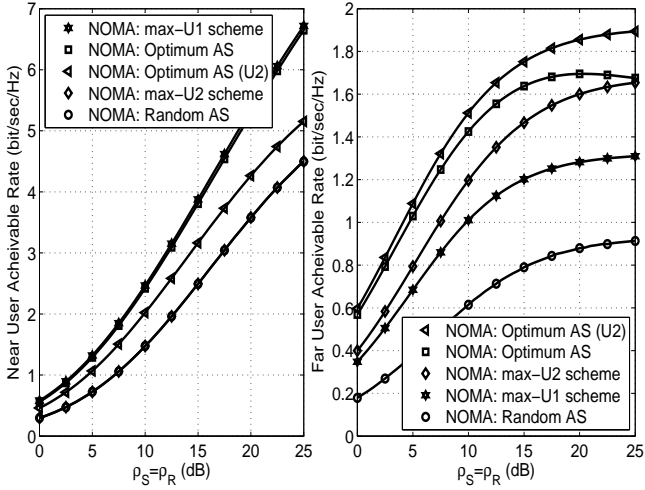}
\vspace{-1.5em}
\caption{ Ergodic rate of near and far users with different AS schemes ($\MB=\MR=\MT=4$, $\Sap=0.3$).}\label{fig:rates}
\vspace{-0.5em}
\end{figure}
%%%%$$$$$$$$$$$$$$$$$$$$$$$$$$$$$$$$$$$$$$$$$$$$$$$$$$$$$$$$$$$$$$$$$$$$$$$$$$$$$$$$$$$$$$$$$$$$$$$$$$$$$$$$$$$$$$$$$$$$$$$$$
\emph{Proof:} An outage event at $\SUu$ occurs when it cannot decode the intended signal for $\SUuu$, or when it can decode it but fails to decode its own signal. Therefore, the outage probability of $\SUu$ can be derived as
%===========================
\vspace{0.5em}
\begin{align}\label{eq:outnear}
\Poutnu&\!=\!1\!-\!\Prob\left(\gamma_{12,\mathsf{AS}}>\!\theta_2,
\gamma_{1,\mathsf{AS}}>\!\theta_1\!\right)\!,\nonumber\\
&  = \Prob\left(\frac{\GSNu}{\GRNu+1} < {\zeta}\right)\nonumber\\
& = F_{\gamma_{1,\mathsf{AS}}}\left(a_1\zeta\right).
\end{align}
%===========================

To this end, the desired result is obtained by evaluating~\eqref{eq:cdf:gam1:S1} and~\eqref{eq:cdf:gam1:S2} at $a_1\zeta$.
$\hspace{12em}\blacksquare$

\begin{proposition}
The outage probability of  $\SUuu$ with max-$\SUu$  and max-$\SUuu$ AS scheme, are
respectively given by
%===========================
\vspace{0.5em}
%\begin{subequations}\label{eq:outS2}
\begin{align}
\PoutfuS&\!=\!1\!-\!\MR e^{-\frac{\theta_2}{\bGRFu}}\!\!\sum_{q=0}^{\MR-1}\!\frac{(-1)^q\binom{\MR-1}{q}e^{-\frac{(q+1)\theta_2}{\bGBR\left(a_2-a_1\theta_2\right)}}}
{(q\!+\!1)\left(1\!+\!\frac{\bGSI}{\bGBR}\frac{(q+1)\theta_2}{ a_2-a_1\theta_2}\right)},
\label{eq:outnuS2}
\end{align}
and
\vspace{0.5em}
\begin{align}
\PoutfuSS&=1-\MT\MB\!\sum_{q=0}^{\MT-1}
\frac{ (-1)^q\binom{\MT-1}{q} e^{-\frac{(q+1)\theta_2}{\bGRFu}}}
{(q+1)},\nonumber\\
&\hspace{1em}\times\sum_{p=0}^{\MB-1}\frac{(-1)^p\binom{\MB-1}{p}e^{-\frac{(p+1)\theta_2}{\bGBR\left(a_2-a_1\theta_2\right)}}}
{(p\!+\!1)\left(1\!+\!\frac{\bGSI}{\MR\bGBR}\frac{(p+1)\theta_2}{(a_2-a_1\theta_2)}\right)}\label{eq:outfuS2}\!.
\end{align}
%\end{subequations}
%===========================
\end{proposition}

\emph{Proof:} An outage event at $\SUuu$ occurs if $\Rly$ fails to decode the intended message to $\SUuu$, or $\Rly$ can decode $\SUuu$ signal but $\SUuu$ fails to decode its message. Therefore, the outage probability of $\SUuu$ can be expressed as
%===========================
\vspace{0.5em}
\begin{align}\label{eq:outfar}
\Poutfu&=1-\Prob\left(\GMRi>\theta_2,
\GRFui>\theta_2\right)\nonumber\\
&=1-\Prob\left(\GMRi>\theta_2\right)\Prob\left(
\GRFui>\theta_2\right).
\end{align}
To this end by substituting the corresponding probabilities presented in subsection~\ref{subsec:rate} into~\eqref{eq:outfar}, we arrive at~\eqref{eq:outnuS2} and~\eqref{eq:outfuS2}.
$\hspace{18.5em}\blacksquare$

%%%%$$$$$$$$$$$$$$$$$$$$$$$$$$$$$$$$$$$$$$$$$$$$$$$$$$$$$$$$$$$$$$$$$$$$$$$$$$$$$$$$$$$$$$$$$$$$$$$$$$$$$$$$$$$$$$$$$$$$$$$$$
\vspace{1em}
\begin{figure}[htb!]
\centering
\includegraphics[width=85mm, height=66mm]{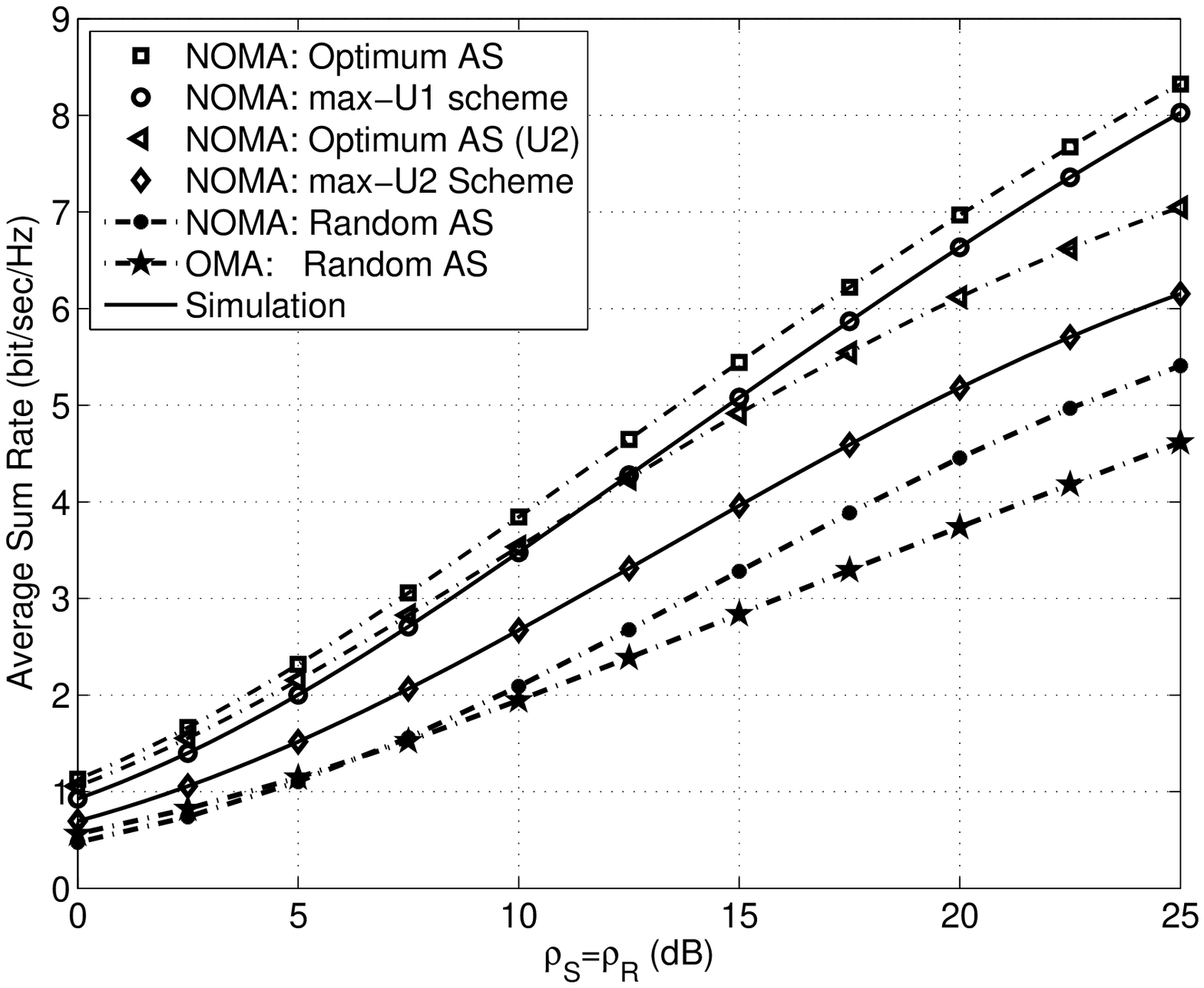}
\vspace{-1.5em}
\caption{ Ergodic sum rate with different AS schemes ($\MB=\MR=\MT=4$, $\Sap=0.3$).}\label{fig:sumrate}
\vspace{-1em}
\end{figure}
%%%%$$$$$$$$$$$$$$$$$$$$$$$$$$$$$$$$$$$$$$$$$$$$$$$$$$$$$$$$$$$$$$$$$$$$$$$$$$$$$$$$$$$$$$$$$$$$$$$$$$$$$$$$$$$$$$$$$$$$$$$$$
%===========================================================
\section{Numerical Results and Discussion} \label{sec:num}
%=============================================================
In this section, we present numerical results to quantify the performance gains when max-$\SUu$ and max-$\SUuu$ AS schemes are adopted in the considered  FD cooperative NOMA system. We set $a_1=0.25$, $a_2=0.75$ and $k_1 = 0.01$.

Fig.~\ref{fig:rates} shows the ergodic rates of $\SUu$ and $\SUuu$ with the proposed AS schemes.  We have also plotted the curves for \emph{i) {Optimum AS scheme}}: that performs an exhaustive search of all possible combinations to determine the antenna subset in order to maximize the ergodic sum rate, \emph{ii) {Optimum AS ($\SUuu$) scheme} }: This scheme aims to maximize the \emph{e2e} SINR at $\SUuu$ in an optimal sense and performs an exhaustive search of all possible combinations to determine the optimum antenna subset in order to maximize the \emph{e2e} SINR at $\SUuu$, and \emph{iii) {Random AS scheme}}: that performs random AS at the BS and relay input/output. It can be observed that the max-$\SUu$ AS scheme is able to improve the ergodic rate of both $\SUu$ and $\SUuu$, while max-$\SUuu$ AS scheme, only increases the ergodic rate of $\SUuu$. The max-$\SUu$ AS scheme provides the best performance for $\SUu$, while  the performances of the max-$\SUuu$  AS scheme and random AS scheme for $\SUu$ are almost identical. However, both max-$\SUu$ AS scheme and max-$\SUuu$ AS scheme are able to increase the ergodic rate of $\SUuu$. Moreover, we see that with increasing transmit power at BS and $\Rly$, max-$\SUuu$ AS scheme achieves the same performance as Optimum AS scheme.

In Fig.~\ref{fig:sumrate} the ergodic sum rate with different AS schemes is illustrated. It is evident that the ergodic achievable  sum rate with the proposed AS schemes is higher than that of OMA FD relay system. In all transmit power regimes the performance gap between the max-$\SUu$ AS scheme and Optimum AS scheme is negligible. Moreover, the difference between the max-$\SUu$ and Optimum AS ($\SUuu$) enlarges when transmit power at the BS and $\Rly$ is increased and remains constant for medium-to-high transmit power values.

%%%%$$$$$$$$$$$$$$$$$$$$$$$$$$$$$$$$$$$$$$$$$$$$$$$$$$$$$$$$$$$$$$$$$$$$$$$$$$$$$$$$$$$$$$$$$$$$$$$$$$$$$$$$$$$$$$$$$$$$$$$$$
\begin{figure}[]
\centering
\includegraphics[width=85mm, height=58mm]{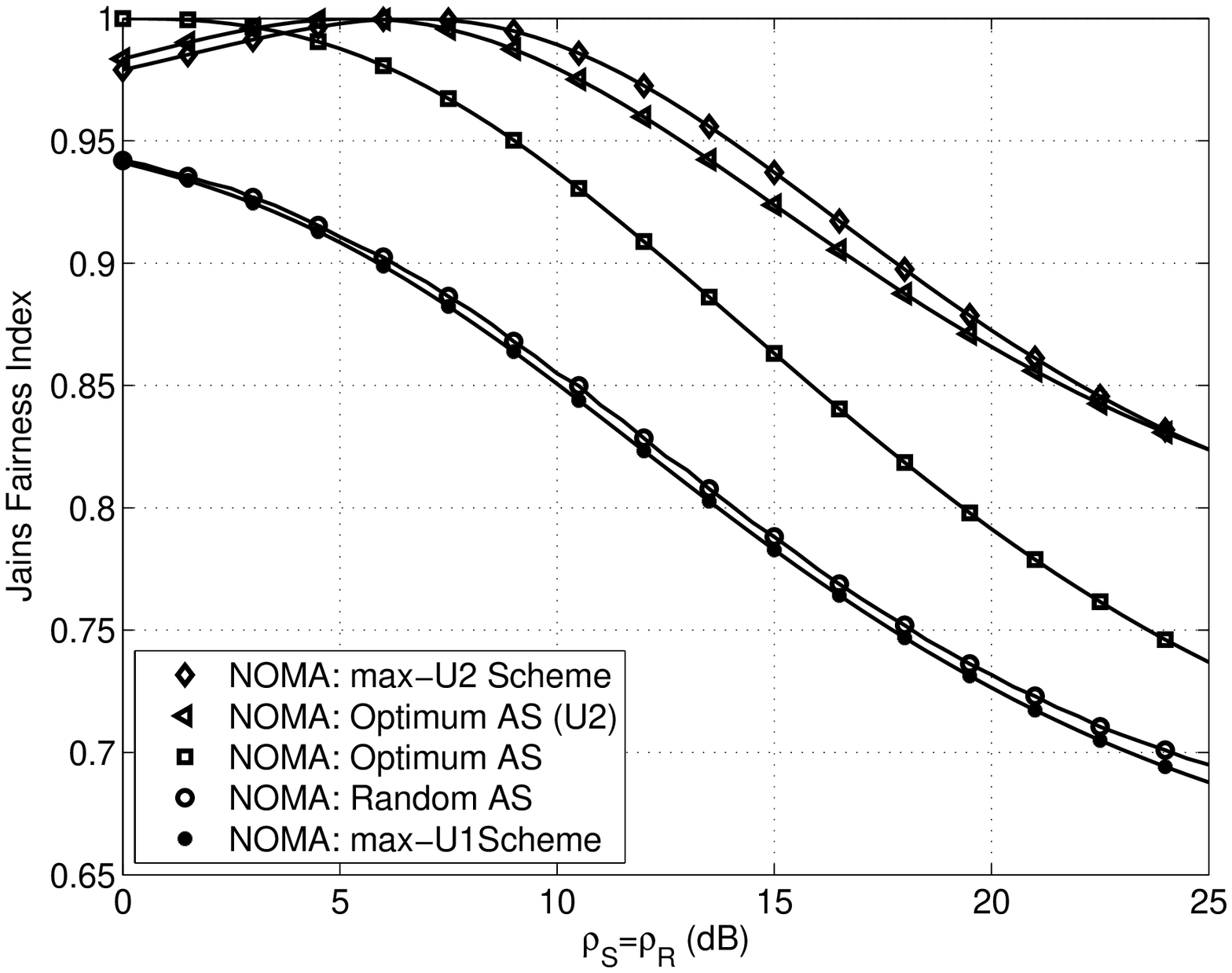}
\vspace{-1.7em}
\caption{ Jain's fairness index versus transmit power ($\MB=\MR=\MT=4$, $\Sap=0.3$).}\label{fig:fairnesindex}
\vspace{-1.2em}
\end{figure}
%%%%$$$$$$$$$$$$$$$$$$$$$$$$$$$$$$$$$$$$$$$$$$$$$$$$$$$$$$$$$$$$$$$$$$$$$$$$$$$$$$$$$$$$$$$$$$$$$$$$$$$$$$$$$$$$$$$$$$$$$$$$$

%%%%$$$$$$$$$$$$$$$$$$$$$$$$$$$$$$$$$$$$$$$$$$$$$$$$$$$$$$$$$$$$$$$$$$$$$$$$$$$$$$$$$$$$$$$$$$$$$$$$$$$$$$$$$$$$$$$$$$$$$$$$$
\begin{figure}[]
\centering
\includegraphics[width=85mm, height=58mm]{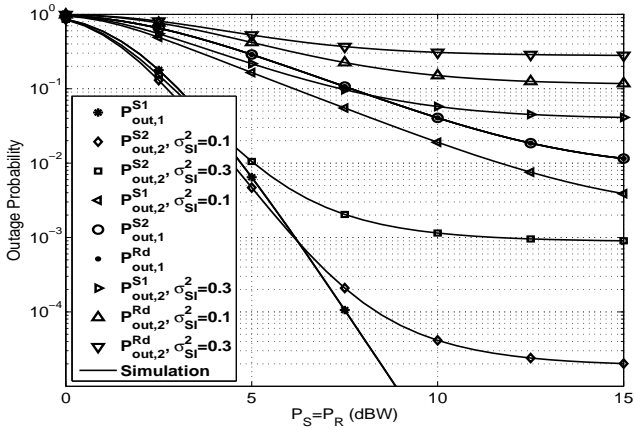}
\vspace{-1.7em}
\caption{ Outage probability versus transmit power ($\MB=\MR=\MT=4$, $\mathcal{R}_1=\mathcal{R}_2 = 0.5$ bps/Hz).}\label{fig:outage}
\vspace{-1.2em}
\end{figure}
%%%%$$$$$$$$$$$$$$$$$$$$$$$$$$$$$$$$$$$$$$$$$$$$$$$$$$$$$$$$$$$$$$$$$$$$$$$$$$$$$$$$$$$$$$$$$$$$$$$$$$$$$$$$$$$$$$$$$$$$$$$$$

Fig.~\ref{fig:fairnesindex} shows the Jain's fairness index versus transmit power and for different AS schemes.  This index
is a bounded continuous function and is the most used quantitative measure to study the fairness in wireless systems. The Jain's fairness index for the considered dual-users scenario can be expressed as~\cite{fairness}
%===================
\vspace{0em}
\begin{align*}
 J = \frac{(\mathcal{R}_{\SUu}^{\mathsf{AS}}+\mathcal{R}_{\SUuu}^{\mathsf{AS}})^2}{2((\mathcal{R}_{\SUu}^{\mathsf{AS}})^2 + (\mathcal{R}_{\SUuu}^{\mathsf{AS}})^2)},
 \end{align*}
 %====================
which its range is the interval $[\frac{1}{2}, 1]$. In this interval, $J=\frac{1}{2}$
corresponds to the least fair allocation in which only one user receives a non-zero rate, and $J=1$ corresponds to the fairest allocation in which both near and far user receive the same rate. From Fig.~\ref{fig:fairnesindex} We see that in the low SNR region, Optimum AS scheme achieves the best user fairness out of all AS schemes. However, max-$\SUuu$ AS scheme can provide a better user fairness in medium-to-high SNR regimes and hence can balance the tradeoff between the ergodic rate and user fairness.

Fig.~\ref{fig:outage} shows the outage probability of the AS schemes with different residual SI strengths. The max-$\SUu$ AS scheme provides the  best outage performance for $\SUu$ and improves the outage performance of $\SUuu$. Moreover, the max-$\SUuu$ AS scheme can significantly improve the outage performance of $\SUuu$, while it exhibits the same performance as Random AS scheme for $\SUu$. Finally, due to inter-user interference at $\SUu$ and SI at $\Rly$ our results show that all AS schemes suffer from zero-order diversity. However, the proposed AS schemes can reduce the error floor.

%==================================================
\vspace{-0.7em}
\section{Conclusion} \label{sec:conc}
%==================================================
In this paper, we have studied the AS problem for a FD cooperative NOMA system. Two low complexity AS schemes, namely max-$\SUu$ AS scheme and max-$\SUuu$ AS scheme were proposed to maximize the \emph{e2e} SINR at the near and far user, respectively. The performance of both AS schemes have been characterized in terms of the ergodic sum rate and outage probability. Our results revealed that the max-$\SUu$ AS scheme achieves near optimum sum-rate performance while the  max-$\SUuu$ AS scheme exhibits a better user fairness. Moreover, the outage performance of the near user can be significantly improved by using the max-$\SUu$ AS scheme, while the outage probability of the far user can be effectively improved via the max-$\SUuu$ AS scheme.

%=======================================
\bibliographystyle{IEEEtran}

%=======================================

\end{document}